\documentstyle[aps,12pt]{revtex}

\begin{document}
\title{The $^{5}D$ term origin of the excited triplet in LaCoO$_{3}$}
\author{Z. Ropka}
\address{Center for Solid State Physics, S$^{nt}$ Filip 5,31-150Krakow,Poland%
}
\author{R. J. Radwanski}
\address{Center for Solid State Physics, S$^{nt}$ Filip 5,31-150Krakow,Poland%
\\
Institute of Physics, Pedagogical University, 30-084Krakow, Poland\\
email: sfradwan@cyf-kr.edu.pl, http://www.css-physics.edu.pl}
\maketitle

\begin{abstract}
We provide the proof for the $^{5}D$ term origin of an excited triplet
observed in the recent electron-spin- resonance (ESR) experiments by Noguchi
et al. (Phys. Rev. B {\bf 66}, 094404 (2002)). We have succeeded to fully
describe experimental ESR results both for the zero-field $g$-factor, of
3.35, and the splitting $D$ of 4.90 cm$^{-1}$, as well as for the magnetic
field applied along different crystallographic directions within the
localized electron atomic-like approach as originating from excitations
within the lowest triplet of the $^{5}T_{2g}$ octahedral subterm of the $%
^{5}D$ term. In our atomic-like approach the {\it d} electrons of the Co$%
^{3+}$ ion in LaCoO$_{3}$ form the highly-correlated atomic-like 3$d^{6}$
system with the singlet $^{1}A_{1}$ ground state (an octahedral subterm of
the $^{1}I$ term) and the excited octahedral $^{5}T_{2g}$ subterm of the $%
^{5}D$ term. We take the ESR\ experiment as confirmation of the existence of
the discrete electronic structure for 3$d$ electron states in LaCoO$_{3}$ on
the meV scale.

PACS\ No: 76.30.Fc; 75.10.Dg : 73.30.-m;

Keywords: electronic structure, crystal field, spin-orbit coupling, LaCoO$%
_{3}$

Submitted 24.12.2002 to Phys. Rev. B.
\end{abstract}

\date{(24.12.2002)}

{\bf 1. Introduction}

LaCoO$_{3}$ attracts much attention in recent 50 years due to its
non-magnetic ground state and the significant violation of the Curie-Weiss
law at low temperatures\cite{1,2} often discussed in terms of successive
changes of spin states with the increasing temperature. A great number of
band-structure calculations \cite{3,4,5,6,7,8,9}, yielding the continuous
energy spectrum for 3$d$ states spread over 6 eV, as well as crystal-field
(CEF) considerations with discrete, in the meV energy scale, energy spectrum %
\cite{11,12,13} did not yield final convincing solution for the magnetism,
the electronic structure and thermodynamics for LaCoO$_{3}$. A short
comparative discussion of different modern theoretical approaches to LaCoO$%
_{3}$ one can find in Ref. \cite{14}.

Recently Noguchi et al. \cite{15} have succeeded in the Electron Spin
Resonance (ESR) measurements for an excited triplet, behavior of which under
external magnetic field applied along different crystallographic directions
of single-crystalline LaCoO$_{3}$ is very well described by an {\bf effective%
} spin-Hamiltonian, Eq. 3.1 of Ref. \cite{15}, with the {\bf effective spin
S=1} and with fitted values $g_{\parallel }$=3.35, $g_{\perp }$=3.55 and $D$%
= +4.90 cm$^{-1}$ (=7.056 K=0.6 meV). This excited triplet has been
discussed in connection to the intermediate-spin state in the standard
description of LaCoO$_{3}$ in terms of low-spin (LS), intermediate-spin (IS)
and high-spin (HS)\ states \cite{4,16}.

The aim of this paper is to provide the proof that the triplet state
revealed in the recent ESR\ experiment of Noguchi et al. is the state
originating from the atomic-like term $^{5}D$ of the 3$d^{6}$ electron
system existing in the Co$^{3+}$ ion. This 25-fold degenerate $^{5}D$ term
(in the $\left| LSL_{\text{z}}S_{\text{z}}\right\rangle $ space) and its
15-fold degenerate $^{5}T_{2g}$ octahedral subterm we have studied in great
details as we have considered its singlet as the ground state in our
up-to-now description of LaCoO$_{3}$, see Fig. 1c of Ref. \cite{12} and Fig.
2c of \cite{13}.

{\bf 2. Theoretical outline}

For the description of the excited triplet we apply exactly the same
Hamiltonian as has been used by us in Refs \cite{11,12,13} namely Eq. 1 of
Ref.\cite{12} and of Ref.\cite{13} for the term $^{5}D$ with $S$=2 and $L$=2:

\begin{equation}
H_{d}=H_{cub}(L,L_{z})+\lambda LS+B_{2}^{0}O\,_{2}^{0}(L,L_{z})+\mu
_{B}(L+g_{s}S)B  \eqnum{1}
\end{equation}

where $H_{cub}=B_{4}\cdot (O\,_{4}^{0}+5O\,_{4}^{4})$ for $z$ axis along the
cube edge and $O\,_{m}^{n}$ are the Stevens operators collected in, e.g. %
\cite{17}, table 16, p. 863. The cubic CEF Hamiltonian takes, for the $z$
axis along the cube diagonal, the form $H_{cub}=-$ $\frac{2}{3}B_{4}\cdot
(O\,_{4}^{0}-20\sqrt{2}O\,_{4}^{3})$. This later form is useful for LaCoO$%
_{3}$ due to the experimentally observed rhombohedral (trigonal) distortion
that can be then described by the parameter B$_{2}^{0}$. For the 3$d^{6}$
system in the oxygen octahedron B$_{4}$%
\mbox{$>$}%
0 \cite{11,12,13,17}, p. 374. The last term in Eq. (1) allows studies of the
influence of the magnetic field B, in the present case used for computer
ESR\ experiments. The energy states, calculated for the dominant octahedral
crystal field, weaker intra-atomic spin-orbit interactions and the weakest
trigonal distortion, are shown in Fig. 1d and Fig. 2c of Ref.\cite{11} and
in Fig. 1d of Ref. \cite{18}. According to us the observed triplet is the
one shown as the lowest triplet in these figures.

{\bf 3. Results and discussion}

The splitting of the triplet, into the singlet and doublet, results from the
trigonal distortion. This splitting, denoted as $D$ in Ref.\cite{15}, is
roughly proportional to the trigonal distortion B$_{2}^{0}$ parameter, Fig.
1. The magnetic moment, equal to the $g$ factor in case of the triplet, of
the doublet decreases with the increasing distortion from a value of 3.53 $%
\mu _{B}$ found for the purely octahedral crystal field in the presence of
the spin-orbit coupling, Fig. 1c of Ref. \cite{12}. Fig. 2 shows variation
of the magnetic moment {\it vs} the splitting $D$ for different values of
the spin-orbit coupling in the presence of the octahedral crystal field
described by the octahedral parameter B$_{4}$=+200 K. The experimentally
observed values $D$=+4.90 cm$^{-1}$ (for $D>0$ the singlet is lower) and $g$
= 3.35 are obtained for $\lambda $ = -185 K and B$_{2}^{0}$=+7.2 K.
Practically the same curves are obtained for larger values of B$_{4}$ - it
is related to the fact that B$_{4}$ determines the splitting of the $^{5}D$
term and the 2-3 eV energy scale.

The calculated dependence of the triplet states under the action of the
external magnetic field applied along different crystallographic directions,
denoted in Fig. 1 of Ref. \cite{15}, is shown on Figs 3, 4 and 5 together
with the ESR transitions f$_{o}$, f$_{1}$ and f$_{2}$ for selected
frequencies. For clarity, we have performed calculations of all 25 states in
the $\left| LSL_{\text{z}}S_{\text{z}}\right\rangle $ space \cite{19}, but
only the three lowest states are shown in Figs 3-5. The energy structure of
the 15 lowest states, originating from the $^{5}T_{2g}$ subterm, is shown in
Fig. 6. It is predominantly determined by the spin-orbit interactions that
define the 0.1 eV energy scale. The comparison of our Fig. 3 with Fig. 4 of
Ref. \cite{15}, Fig. 4 with Fig. 5a of Ref. \cite{15} and Fig. 5 with Fig.
5b of Ref. \cite{15} reveals a perfect reproduction of the experimental ESR\
results. In particular, we would like to put attention to the reproduction
of the behavior for the cubic-edge direction (Fig. 4) and for the direction
perpendicular to the cube diagonal (Fig. 5). For the cubic direction the
resonance fields f$_{1}$ and f$_{2}$ are the same what is not the case for
the perpendicular direction. For the [001] direction, the cube diagonal, the
resonance fields are shifted by 6 T, Fig. 3, in agreement with the
experimental observation, shown in Fig. 4 of Ref. \cite{15}.

It is worth to point out that for the shown description we use only 3
parameters: B$_{4}$ (not so sensitive provided that it is positive and it is
in the range 50-300 K), $\lambda $ and B$_{2}^{0}$, that have been selected
as shown in Figs 1 and 2. All of them have clear physical meaning. In fact,
the most important is the choice of the atomic term, the term $^{5}D$ with $%
S $=2 and $L$=2 in the present case. In this choice we have been oriented by
our assumption of the fulfilling of two Hund's rules also in case of solids.
The spin-orbit coupling parameter is by 12\% smaller than the free-ion
value, what is reasonable. But in fact, the most important ''unwritten''
assumption is an assumption about the significant surviving of the atomic
states of the 3$d^{6}$ system - it corresponds to the very strong
correlations among 3$d$ electrons. Thanks these strong intra-atomic
correlations the atomic-like terms of the Co$^{3+}$ ion survive also in the
solid when this cation becomes the full part of a solid.

{\bf 4. Conclusions and remarks}

We proved that the excited triplet, revealed in the ESR experiment by
Noguchi et al. \cite{15}, originates from the $^{5}D$ term, more exactly
from the lowest triplet of its $^{5}T_{2g}$ subterm. The zero-field $g$%
-factor and the singlet-doublet splitting as well as the ESR\ excitations
for different crystallographic directions have been very well described by
taking into account the octahedral CEF interactions, the intra-atomic
spin-orbit coupling and a relatively weak trigonal distortion. So good and
very consistent description with so limited number of parameters provides
significantly large confidence to the found physical origin of the excited
triplet. It is worth to add that the same approach we have used for FeBr$%
_{2} $ and the term $^{5}D$ has been found to be the ground term of the Fe$%
^{2+}$ ion \cite{18}. In FeBr$_{2}$ the value of $D$ was derived to be -33 K
and also has resulted from the trigonal distortion B$_{2}^{0}$ of -30 K. The
splitting $D$ in both compounds is of the same size in the absolute value.
However, in LaCoO$_{3}$ the $^{5}D$ term is the excited term as thanks the
activated character of the ESR\ spectra we became sure that there is an
extra singlet at $\Delta _{s-t}$ =140 K below - as is shown in Fig. 6c. This
singlet it is a state $^{1}A_{1}$. It originates from the $^{1}I$ term, that
in the free Co$^{3+}$ (CoIV) ion lies 4.43 eV above the ground term \cite{20}%
. The 13-fold degenerate term $^{1}I$ splits under the action of the
octahedral crystal field into 6 states and the singlet $^{1}A_{1}$ state is
strongly pushed down due to its very large value of the orbital quantum
number $L$=6. The situation with the $^{1}A_{1}$ subterm being slightly
lower than the $^{5}T_{2g}$ subterm occurs for $Dq$/$B$=2.025 on the
Tanabe-Sugano diagrams presented on p. 388 of Ref. \cite{17}. Assuming the
Racah parameter $B$ of 1065 cm$^{-1}$ one gets $Dq$=2156 cm$^{-1}$ and
consequently B$_{4}$=+260 K. This value is only slightly larger, by 30\%,
than we used previously \cite{11,12,13} and in the calculations of ESR
dependences on Figs 3-5. Thus, the crystal field interactions are strong in
LaCoO$_{3}$ but not so strong to destroy the atomic discrete electronic
structure of the 3$d$-electron states as is postulated by the Quantum
Atomistic Solid-State theory (QUASST) \cite{21,22,23}. The inferred
electronic structure, in the 4 eV scale, of cubic subterms of the Co$^{3+}$
ion in the octahedral crystal field, relevant to LaCoO$_{3}$, is shown in
Fig. 7. Surely, our long-lasting studies as well as growing number of more
and more sophisticated experiments indicate that it is the highest time to
''unquench'' the orbital moment in the solid-state physics for description
of the magnetism and the electronic structure of 3$d$-atom containing
compounds \cite{21,22,23}.

Acknowledgement. The authors are grateful to M. Kocor and A.J. Baran for the
assistance in the computer preparation of this paper.

Fig. 1. Calculated dependence of the splitting $D$ of the lowest triplet,
into a singlet and excited doublet, of the $^{5}T_{2g}$ subterm on the
distortion trigonal parameter B$_{2}^{0}$ in the presence of the octahedral
crystal field B$_{4}$=+200 K and the spin-orbit coupling $\lambda $=-185 K.

Fig. 2. Calculated dependence of the magnetic moment of the excited doublet
on the splitting $D$ for different values of the spin-orbit coupling $%
\lambda $ in the presence of the octahedral crystal field B$_{4}$=+200 K.
Values of $g_{\parallel }$=3.35 and $D$= +4.90 cm$^{-1}$ (=7.056 K=0.6 meV)
are reproduced by $\lambda $=-185 K and the trigonal distortion parameter B$%
_{2}^{0}$=+7.2 K. Curves for a larger value of B$_{4}$=+260 K are only
slightly above (less than 1\%).

Fig. 3. Calculated field dependence of the three lowest states, the quasi
triplet, of the 25-fold degenerate $^{5}D$ term for the octahedral
crystal-field parameter B$_{4}$=+200 K, the spin-orbit coupling $\lambda $%
=-185 K and the distortion trigonal parameter B$_{2}^{0}$=+7.2 K for
external magnetic fields applied along the diagonal of the cube of the
perovskite structure of LaCoO$_{3}$. The ESR\ transitions f$_{1}$ and f$_{2}$
for the frequency of 1000 GHz (=48 K) are shown. The zero energy is at the
level of the unsplit $^{5}D$ term.

Fig. 4. Calculated field dependence of the three lowest states of the $^{5}D$
term for the octahedral crystal field B$_{4}$=+200 K, the spin-orbit
coupling $\lambda $=-185 K and the distortion trigonal parameter B$_{2}^{0}$%
=+7.2 K for external magnetic fields applied along the edge of the cube of
the perovskite structure of LaCoO$_{3}$. The ESR\ transitions f$_{o}$, f$%
_{1} $ and f$_{2}$ for the frequency of 760 GHz (=36.5 K) are shown. The
ESR\ transitions f$_{1}$ and f$_{2}$ occur practically at the same magnetic
field, what is a characteristic feature for the cubic direction.

Fig. 5. Calculated field dependence of the three lowest states of the $^{5}D$
term for the octahedral crystal field B$_{4}$=+200 K, the spin-orbit
coupling $\lambda $=-185 K and the distortion trigonal parameter B$_{2}^{0}$%
=+7.2 K for external magnetic fields applied along the [110] direction, i.e.
perpendicular to the diagonal of the cube of the perovskite structure of
LaCoO$_{3}$. The ESR\ transitions f$_{o}$, f$_{1}$ and f$_{2}$ for the
frequency of 429 GHz (=20.6 K) are shown.

Fig. 6. Calculated low-energy electronic structure of the Co$^{3+}$ ion in
LaCoO$_{3}$ originating from the $^{5}T_{2g}$ cubic subterm with the $%
^{1}A_{1}$ singlet ground subterm put 140 K below the lowest $^{5}T_{2g}$
state. Such the structure is produced by the dominant octahedral
crystal-field interactions and the intra-atomic spin-orbit coupling (b). c)
shows the splitting produced by the trigonal distortion. The states are
labelled by the degeneracy, the magnetic moment and the energy with respect
to the lowest state of the $^{5}D$ term.

Fig. 7. Electronic structure of cubic subterms of the Co$^{3+}$ ion in the
octahedral crystal field inferred from the Tanabe-Sugano diagram for $Dq$/$B$
=2.025 relevant to LaCoO$_{3}$. The arrow indicates the strength of the
octahedral crystal field considered by us in the previous studies. The
crystal field is slightly stronger but still not so strong to destroy the
atomic-like discrete structure of the $d$-electron states as is postulated
by the Quantum Atomistic Solid-State theory.

\end{document}